# High $Q$ micro-ring resonators fabricated from polycrystalline aluminum nitride films for near infrared and visible photonics

**Wolfram H.P. Pernice,[†] Chi Xiong, and Hong X. Tang[*]**

*Department of Electrical Engineering, Yale University, New Haven, CT 06511, USA*
[†]*Current address: Institute of Nanotechnology, Karlsruhe Institute of Technology, 76344 Eggenstein-Leopoldshafen, Germany*
[*]*hong.tang@yale.edu*

**Abstract:** We demonstrate wideband integrated photonic circuits in sputter-deposited aluminum nitride (AlN) thin films. At both near-infrared and visible wavelengths, we achieve low propagation loss in integrated waveguides and realize high-quality optical resonators. In the telecoms C-band (1520-1580 nm), we obtain the highest optical $Q$ factor of 440,000. Critical coupled devices show extinction ratio above 30 dB. For visible wavelengths (around 770 nm), intrinsic quality factors in excess of 30,000 is demonstrated. Our work illustrates the potential of AlN as a low loss material for wideband optical applications.

## 1. Introduction

The commercial excitement for nanophotonics is based on the perceived advantages of large-scale, monolithic integration of photonic circuits with electronic circuits [1]. Since the realization of silicon-on-insulator wafers, densely integrated photonic circuits fabricated from silicon thin films have demonstrated excellent performance in a range of applications [2-4]. Due to the relatively small bandgap of silicon (1.1 eV [5]), silicon waveguides are however restricted to operation wavelengths above 1100 nm. For wider band applications covering visible wavelengths, photonic devices have to be realized in different material systems that are compatible with silicon fabrication processes. Within the group of wide bandgap semiconductors, silicon nitride ($Si_3N_4$) has attracted particular interest for the realization of CMOS compatible nanophotonic devices [6-10]. Low material absorption [11,12] and good mechanical properties [13,14] make $Si_3N_4$ also a prime candidate for optomechanical devices. However, the deposition of thick $Si_3N_4$ films of high quality, which are required for nanophotonic waveguides at longer wavelengths, is often accompanied by large internal stress, which leads to film cracking.

Here we present a new material system for integrated optics based on aluminum nitride (AlN) thin films. Because of its large piezoelectric coefficient, AlN has been widely utilized in high-performance electromechanical devices including surface acoustic wave (SAW) devices [15], thin-film bulk acoustic wave resonators (FBARs) [16], contour-mode resonators [17], and lamb wave resonators [18]. Moreover, AlN is a wide-band semiconductor with a transmission window spanning from 200 nm to 13.6 μm [19], due the large bandgap of 6.2 eV [20]. Wafer-scale substrates with good uniformity can be prepared by sputter deposition, thus allowing for large-scale photonic designs. Sputter-deposited AlN films do not show an in-plane crystallographic preference orientation and therefore etching anisotropy as encountered in crystalline AlN films is not observed in our experiments. Despite of the existence of grain boundaries in similarly prepared films, we obtain low propagation loss of 0.8 dB/cm in nanophotonic waveguides at near-infrared (NIR) wavelengths, which allows us to realize high-*Q* optical micro-ring cavities. We study the coupling properties of ring resonators with different waveguide widths and find best optical *Q* of 440,000. In addition, we fabricate and measure ring resonators working in the visible regime (766 nm-780 nm) with efficient on- and off-chip coupling through grating couplers. The near-intrinsic quality factors measured around 770 nm exceed 30,000, thus illustrating the potential of AlN for wideband optical applications.

## 2. Substrate preparation and device fabrication

To fabricate the devices used in the current study, we first develop AlN-on-insulator substrates with a layer structure that resembles the silicon-on-insulator wafers that are widely used in silicon photonics. Silicon carrier wafers are thermally oxidized to provide an optical buffer layer with low refractive index (1.45 for $SiO_2$). The buffer layer is grown to a thickness of 2.6 μm. Subsequently AlN thin films are deposited onto the oxide using a sputtering procedure. The AlN film is highly *c*-axis oriented with a rocking curve full width at half maximum (FWHM) of 2° at AlN's [0002] peak. The film thickness is chosen to be 330 nm similar to previously reported structures in silicon nitride [21]. Since AlN provides a comparable refractive index of 2.1 to silicon nitride's refractive index range (2.0-2.2, depending on the internal stress), photonic devices of similar dimensions can be fabricated.

Optical circuits are patterned using electron beam lithography on a Vistec EBPG 5000+ 100kV writer. Following lithography in hydrogen silsesquioxane (HSQ) resist, the exposed structures are transferred into the AlN thin film using inductively coupled plasma (ICP) reactive ion etching (RIE) in $Cl_2/BCl_3/Ar$ chemistry, a process similar to the GaN fabrication procedure reported previously [22]. In order to prevent oxidization of the AlN film during the RIE process (which leads to formation of etch-resistant alumina layers) the addition of $BCl_3$ is necessary to maintain high etch rates. By adjusting the ICP and RIE parameters we preserve a target etch rate on the order of 200 nm/min, which leads to acceptable surface roughness and sufficient etch selectivity against the HSQ resist.

## 3. Design of grating couplers for visible and IR wavelengths

In order to optically analyze fabricated photonic circuitry we employ focusing grating couplers similar in design to our silicon nitride grating couplers [13]. Light is coupled into and out of the waveguides by first order Bragg scattering on the on-chip grating. A particular coupling wavelength can be selected by adjusting the period of the grating [23], which is optimized by finite-difference time-domain (FDTD) simulations.

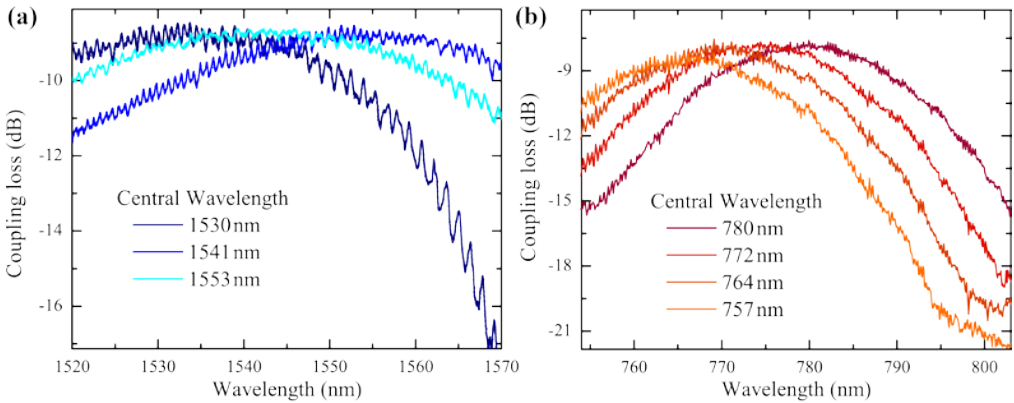

Fig.1 (a) Measured transmission spectra of grating couplers optimized for IR light around 1550nm showing ~60 nm bandwidth. By varying the grating period the central coupling wavelength is selected. (b) Transmission of grating couplers optimized for coupling at 780 nm with a coupling bandwidth of ~30 nm.

The grating period required for central coupling around 770 nm is roughly reduced by half compared to 1550 nm input light. A pair of grating couplers connected by a low-loss nanophotonic waveguide is used to measure the overall transmission through the device.

Because the couplers are closely spaced on the chip and obey optical reciprocity it is assumed that they perform similarly after fabrication, which is also confirmed by reverse transmission measurements. The grating couplers are analyzed by measuring transmission from the input coupler to the output coupler. Around 1550 nm we use a tunable diode laser source to scan across the coupling bandwidth. At 770 nm we employ a superluminescent diode and record the transmitted light with an optical spectrum analyzer. Data shown in Fig. 1(a) and Fig. 1(b) are from couplers optimized for coupling around 1550 nm input and 770 nm input, respectively. The coupling profile is enveloped by Fabry-Perot fringes due to back-reflection between the grating couplers. Because the AlN waveguides have a lower effective refractive index than silicon, the grating coupler is better matched to the fiber modes. As a result, the available coupling bandwidth around 1550 nm is larger, spanning roughly 60 nm. At 770 nm the coupling bandwidth is reduced to about 30 nm. The coupling bandwidth scales with the wavelength of the central coupling point and the refractive index contrast. Because the effective indices of the waveguides at visible and IR wavelengths are comparable, the bandwidth is reduced at shorter wavelengths. Typical insertion loss for grating couplers at 1550 nm is measured to be 8.5±1 dB, while the couplers at 770 nm show slightly better coupling efficiency of 7.7±1 dB at the fundamental mode.

## 4. Design of the pulley-waveguide structure for dual band critical coupling

The grating coupler structure described above is used to probe optical microring resonators coupled to on-chip waveguides. In order to increase the coupling of the collection waveguide to the ring for the visible light operation, the waveguide is wrapped around the ring using a pulley structure [24] as shown in Fig.2.

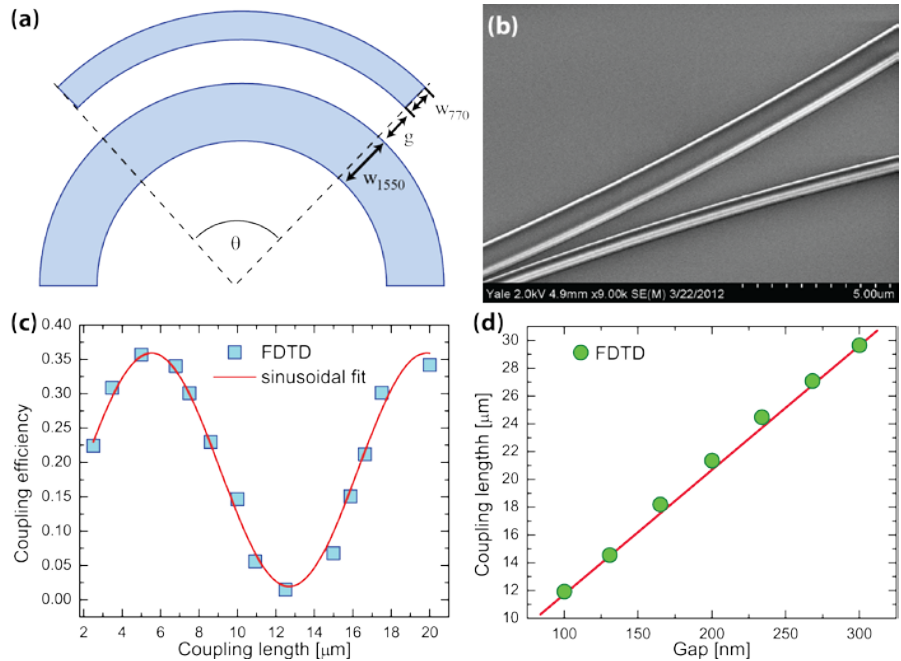

Fig.2. (a) The geometry of the pulley structure used during the FDTD optimization of the coupling gap. Optimization parameters are the coupling gap g and the output waveguide width $w_{770}$. The width of the ring resonator is kept constant. (b) A SEM image of the coupling region of a pulley structure. (c) The calculated power transfer from the ring into the waveguide in dependence of the coupling length at fixed coupling gap. The blue markers represent FDTD simulation results whereas the red line is a sinusoidal fit to the data. (d) The calculated coupling length in dependence of coupling gap. Markers are FDTD results and the red line is the linear fit.

Critical coupling to microring resonators requires miniature separation between the output waveguide and the ring when single-point coupling is chosen. This is restrictive in particular when the coupling wavelength is short as in the case of visible light excitation. The pulley geometry allows for relaxing the requirements on the coupling gap by increasing the coupling length. Results for the AlN geometry are shown in Fig.2. The optimization parameters as shown in Fig. 2(a) are the gap $g$ which separates the pulley waveguide from the ring resonator and the width $w_{770}$ of the pulley. The width $w_{1550}$ of the ring resonator is kept fixed. The optimized parameters are later used during fabrication, as shown in the SEM image of a final device in Fig. 2(b). We employ finite-difference time-domain (FDTD) simulations to calculate the amount of power transferred from the ring into the waveguide in dependence of coupling length, or equivalently, the coverage angle $\theta$. Representative results are shown in Fig. 2(c), where we show the dependence of the out-coupled power on the waveguide separation. We perform a series of calculations with varying coupling length for a given coupling gap. In Fig. 2(c) we show how the power transfer evolves with coupling length (markers: FDTD). From the fit with a sinusoid we can obtain the coupling length for any coupling gap. This is shown in Fig. 2(d), where we plot the fitted coupling length in dependence of waveguide separation. Thus by adjusting the distance between pulley waveguide and ring resonator we can relax the requirements for electron-beam fabrication and work with wider separations, thus avoiding proximity effects during exposure.

## 5. Resonators design and measurement at telecom wavelengths

Fabricated optical ring resonators working around 1550 nm and 770 nm are shown in Fig. 3(a) and Fig. 3(b), respectively. The resonators are coupled to nanophotonic waveguides to allow for the optical characterization in transmission measurements.

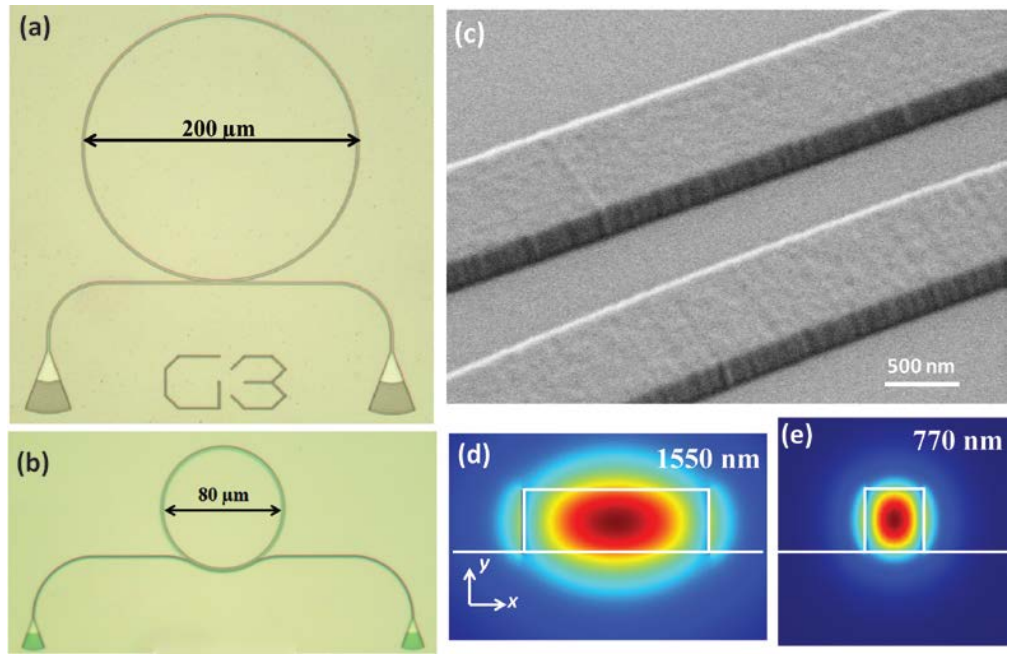

Fig.3 Optical micrographs of fabricated ring resonators for measuring the optical $Q$ of AlN resonators at NIR (a) and visible (b) wavelengths. The 1550 nm input waveguide is 1 µm wide and the 770 nm input waveguide is 350 nm wide. In order to increase the coupling into the ring at visible wavelengths a pulley structure is used. (c) A SEM image of the coupling region of the 1550 nm input into the ring, showing the etch profile of the waveguide sidewalls. Fundamental TE-like mode profile (plot in electric field, x component) in (d) a 1550 nm waveguide (H=330 nm, W=1µm) and (e) a 770 nm waveguide (H=330 nm, W=350 nm).

In Fig. 3(c) we show a SEM image of the coupling region at the 1550 nm resonator. The image reveals a well-defined waveguide structure with relatively low surface roughness. In order to characterize the remaining roughness we perform atomic force microscope (AFM) measurements. We scan the AFM cantilever both across the waveguide surface and along the waveguide sidewall. The AFM data is then analyzed to extract the root-mean-square (rms) roughness. From the obtained data we determine an average top surface roughness of 1.2 nm rms (due to the sputtering process employed) and a sidewall roughness of 3.5 nm. Because both values are well below the target wavelengths, scattering loss is strongly reduced compared to our previously reported results in GaN [22].

Focusing grating couplers are used to couple light into and out of the chip [23]. Two sets of couplers are optimized for IR and visible operation by adjusting the period of the grating. The couplers focus light into input waveguides, which allow for transmission measurements at both 1550 nm and 770 nm. The fundamental TE-like mode profiles for both wavelengths are numerically calculated and shown in Fig. 1(d) and (e). For the visible light operation, in order to increase the coupling of the collection waveguide to the ring, the waveguide is wrapped around the ring using a pulley structure [24]. In order to characterize the fabricated devices we employ the measurement setup shown in Fig. 4(a).

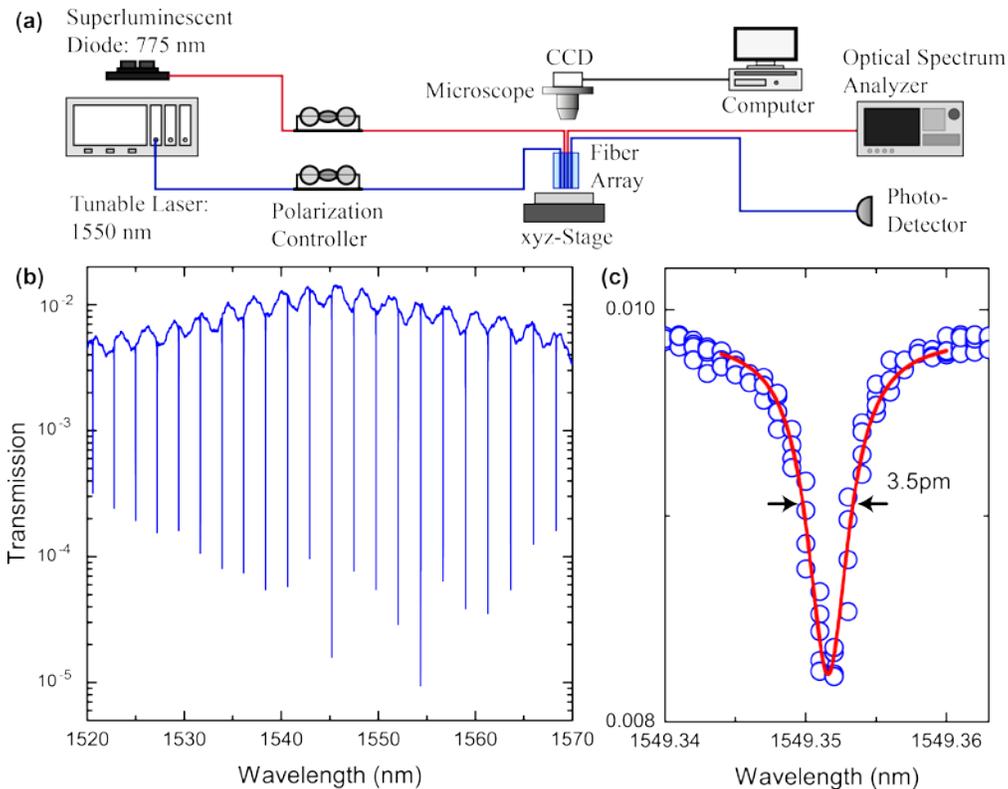

Fig.4 (a) The measurement setup used to characterize the devices. Light from tunable laser sources is coupled into the chip using an optical fiber array. Transmitted NIR and visible light is detected with an InGaAs photoreceiver and a Si photodetector, respectively. (b) The optical transmission spectrum of a critically coupled ring resonator with high extinction ratio of 30 dB for 1550 nm light input. In this case the measured optical $Q$ is on the order of 100,000. (c) The spectrum for an undercoupled ring resonator. The Lorentzian fit to the resonance dip (solid red line) reveals a linewidth of 3.5 pm corresponding to optical $Q$ of 440,000.

Light from two different tunable laser sources New Focus 6428 (NIR) and 6712 (Red) is guided to an optical fiber array that comprises two pairs of single mode fibers for 1550 nm and 770 nm, respectively. Incoming light is adjusted with polarization controllers and collected again at the output port where it is detected by a low-noise InGaAs photoreceiver (New Focus 2011) for the 1550 nm light and a Si photodetector (New Focus 2107) for the 770 nm light. We analyze series of ring resonators with a diameter of 200 µm (Fig. 3(a)) to characterize the optical quality factor at telecom wavelengths. In order to study the coupling behavior of the resonators we fabricate identical devices with coupling gap varying from 200 nm to 1000 nm, thus allowing us to examine ring resonators in both the over-coupled and weakly coupled regimes [25]. The transmission measurement of the device at 1550 nm input shows optical resonances with a free spectral range of 1.9 nm, corresponding to a group index of 2.01. When the coupling gap is optimized for critical coupling we find high extinction ratio of up to 30 dB as shown in Fig. 4(b). In this case the coupling gap amounts to 650 nm and we find optical quality factors on the order of 100,000 from a Lorentzian fit to the resonance. When the coupling gap is further increased the ring is operated in a weakly coupled regime, leading to improved optical $Q$-values which are closer to the resonator's intrinsic quality factor. For devices with a coupling gap of 1000 nm as shown in Fig. 4(c) we measure a cavity linewidth of 3.5 pm, corresponding to high optical $Q$ of 440,000. In the zoom-in spectrum we do not observe resonance splitting of the high-$Q$ mode. Resonance splitting due to coupling between clockwise- and counterclock-wise modes may be revealed if devices with even higher Q can be fabricated. From the fit to the resonance dip the propagation loss of the nanophotonic waveguides can be estimated [26]. Using the expression $\alpha = 10\log_{10} e \cdot 2\pi n_g / Q_{int} \lambda$ (where $Q_{int}$ is the intrinsic quality factor, $\lambda$ the wavelength and $n_g$ is the group index), we determine a propagation loss of α = 0.8 dB/cm, which is on par with state-of-the-art silicon nitride photonic devices. Comparable loss is also found by measuring the attenuation using nanophotonic waveguides with different lengths.

To further characterize the performance of the ring resonators we measure the dependence of the optical quality factor and the extinction ratio on the waveguide width as shown in Fig.5. Single mode waveguiding at 1550 nm is possible in our layer structure down to a simulated cutoff waveguide width of 630 nm.

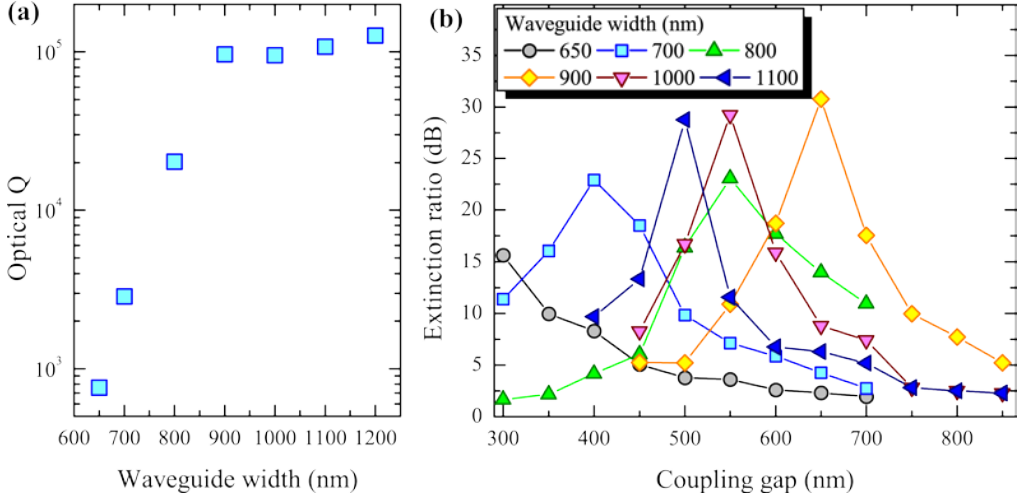

Fig.5 (a) The measured optical quality factor for critically coupled resonator devices in dependence of waveguide width. A monotonic increase in $Q$ is observed for waveguide widths below 900 nm, which is a result of the waveguide width (630 nm) being closer to the cutoff width for 1550 nm input light. (b) The measured extinction ratio in dependence of coupling gap and waveguide width. For the 900 nm wide waveguide the critical coupling gap is the largest (650 nm), indicating that internal losses in the ring are minimized.

However, when the waveguide width of the ring resonator is close to the cutoff bending losses occur and the optical $Q$ is degraded. This is illustrated in Fig. 5(a), where we show the measured optical $Q$ for critically coupled devices with different waveguide width. For each width a series of devices is analyzed with increasing coupling gap. The plotted optical $Q$ is obtained at the point of highest extinction ratio, which is close to the critical coupling condition. A monotonic increase in $Q$ is observed up to a width of 900 nm, where the $Q$ saturates around 100,000. This trend is also observed in terms of maximal extinction when considering the coupling gap required for critical coupling, as shown in Fig. 5(b). Here the measured extinction ratio of fabricated devices is plotted in dependence of waveguide width and coupling gap. In the lower waveguide width range the $Q$ factors are limited because the waveguide width is close to the cutoff wavelength and further by the bend loss due to more extended modes. The plateau around 100,000 in optical $Q$ is a result of the propagation loss inside the ring resonator due to surface roughness scattering. Therefore the optical $Q$ does not further increase once the bend and cutoff losses are not the dominant factors any more.

For narrow waveguides a small coupling gap is required, which indicates that coupling between the feeding waveguide and the ring needs to be strong. Under critical coupling conditions the round-trip loss corresponds to the coupling loss, thus confirming that for narrow waveguides larger internal loss occurs. The widest coupling gap of 650 nm is observed for the 900 nm wide waveguide, suggesting that in this case internal loss is minimized.

**6. Measurement at visible wavelengths**

In addition to characterization at telecom wavelengths, we further perform transmission measurements on the ring resonator shown in Fig. 3(b) in the wavelength range around 770 nm. The propagation loss for a single mode visible waveguide is expected to increase compared with waveguides working at 1550 nm due to more pronounced scattering loss. Here, we design our visible ring resonator to have a radius of 40 µm and a waveguide width of 1 µm to allow for direct comparison with the NIR results. The radius of the ring resonators is chosen to provide more closely matched free-spectral ranges for both IR and visible light. Because the IR resonators have an $FSR_{IR}$ of 1.9 nm a visible ring resonator would have a much smaller $FSR_{vis}$ of $FSR_{IR}/4$ of less than 500 pm. In order to prevent the overlap of resonances in resonators with lower optical $Q$ we therefore employ smaller resonators for shorter wavelengths. The 770 nm feeding waveguide, kept at a narrow width of 350 nm, is wrapped around the ring resonator over a length of 25 μm to facilitate efficient coupling into the ring resonator.

We vary the coupling gap between the pulley structure and the ring from 100 nm to 350 nm. For the smallest gap of 100 nm no optical resonances can be observed in the output. In this case the ring resonator is excited in the overcoupled regime. When the gap is increased to 250 nm, the resonator is near-critically coupled (Fig. 6(a)). In the spectrum we find optical resonances with an extinction ratio near 10 dB and the fitted quality factor of 18,000. The resonances are separated by a FSR of 1.1 nm, corresponding to a group index of 2.2. When the coupling gap is increased further to 300 nm (Fig. 6(b)), the resonator is operated in the undercoupled regime and both the linewidth of the resonances and the extinction ratio are reduced. A quality factor of 30,000, which is expected to be closer to its intrinsic value, is calculated from the fitted linewidth of 26 pm. From the resonance fit we calculate a propagation loss at visible wavelengths of 20 dB/cm. The increase in propagation loss is expected, given that the residual surface roughness provides more significant Rayleigh scattering ($\sim 1/\lambda^4$) at shorter wavelengths. Furthermore, both the refractive index and the material absorption of AlN increase with decreasing wavelength. As a result, the optical mode experiences a stronger refractive index contrast and is thus more susceptible to roughness scattering. The combined effect leads to enhanced propagation loss and thus reduced optical quality factors. Improved etching procedures may be employed to smooth the waveguide sidewalls in future experiments.

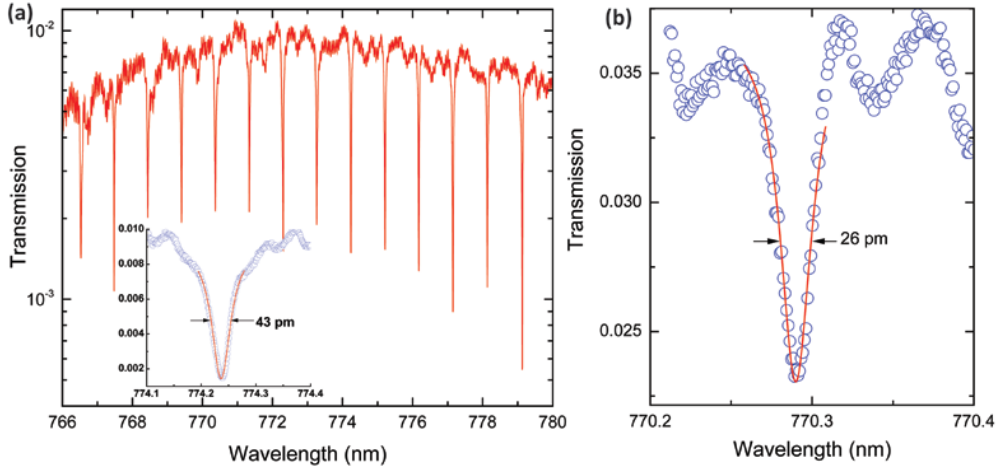

Fig.6 (a) The TE-polarized transmission as a function of wavelength for a ring resonator with R=40 µm. Inset: resonance at 774.24 nm. The fitted linewidth (red line) is 43 pm, corresponding to a *Q* of 18,000. (b) The zoom-in spectrum of an undercoupled ring resonator. The fitted linewidth (red line) is 26 pm, corresponding to a *Q* of 30,000.

**7. Conclusions**

Due to its very wide transparency window, spanning from the mid-IR all the way to visible/UV wavelengths (220 nm~13.6 µm), AlN is a promising candidate for integrated optical applications. Since AlN thin films can be sputter-deposited onto suitable substrates, wafer-scale fabrication is feasible. As shown here, high quality photonic components can be fabricated from AlN with standard nanofabrication routines and are thus compatible with traditional CMOS manufacturing. The potential for seamless integration on silicon substrates using our AlN-on-insulator approach provides a viable route towards a multitude of wideband optical applications. Low propagation loss makes it possible to generate large-area photonic circuits for applications in nonlinear and traditional photonic devices.

**Acknowledgements**

This work was supported by NSF and Packard Foundation. W.H.P. Pernice acknowledges support by DFG grant PE 1832/1-1. C.X. and H.X.T. acknowledges support from the NSF grant through MRSEC DMR 1119826. The authors wish to thank Dr. Michael Rooks, Michael Power, James Agresta and Christopher Tillinghast for assistance in device fabrication.